Research **Open Access**

# Functional architecture of *Escherichia coli*: new insights provided by a natural decomposition approach

Julio A Freyre-González, José A Alonso-Pavón, Luis G Treviño-Quintanilla and Julio Collado-Vides

Address: Programa de Genómica Computacional, Centro de Ciencias Genómicas, Universidad Nacional Autónoma de México. Av. Universidad s/n, Col. Chamilpa 62210, Cuernavaca, Morelos, México.

Correspondence: Julio A Freyre-González. Email: jfreyre@ccg.unam.mx. Julio Collado-Vides. Email: collado@ccg.unam.mx

## Abstract

**Background:** Previous studies have used different methods in an effort to extract the modular organization of transcriptional regulatory networks. However, these approaches are not natural, as they try to cluster strongly connected genes into a module or locate known pleiotropic transcription factors in lower hierarchical layers. Here, we unravel the transcriptional regulatory network of *Escherichia coli* by separating it into its key elements, thus revealing its natural organization. We also present a mathematical criterion, based on the topological features of the transcriptional regulatory network, to classify the network elements into one of two possible classes: hierarchical or modular genes.

**Results:** We found that modular genes are clustered into physiologically correlated groups validated by a statistical analysis of the enrichment of the functional classes. Hierarchical genes encode transcription factors responsible for coordinating module responses based on general interest signals. Hierarchical elements correlate highly with the previously studied global regulators, suggesting that this could be the first mathematical method to identify global regulators. We identified a new element in transcriptional regulatory networks never described before: intermodular genes. These are structural genes that integrate, at the promoter level, signals coming from different modules, and therefore from different physiological responses. Using the concept of pleiotropy, we have reconstructed the hierarchy of the network and discuss the role of feedforward motifs in shaping the hierarchical backbone of the transcriptional regulatory network.

**Conclusions:** This study sheds new light on the design principles underpinning the organization of transcriptional regulatory networks, showing a novel nonpyramidal architecture composed of independent modules globally governed by hierarchical transcription factors, whose responses are integrated by intermodular genes.

## Background

Our understanding of transcriptional control has progressed a long way since Jacob and Monod unraveled the mechanisms that control protein synthesis [1]. These mechanisms allow bacteria to be robust and able to respond to a changing environment. In fact, these regulatory interactions give rise to complex networks [2], which obey organizational principles defining their dynamic behavior [3]. The understanding of these principles is currently a challenge. It has been suggested that decision-making networks require specific topologies [4]. Indeed, there are strong arguments supporting the notion of a modular organization in the cell [5]. A module is defined as a group of cooperating elements with one specific cellular function [2,5]. In genetic networks, these modules must comprise genes that respond in a coordinated way under the influence of specific stimuli [5-7].

Topological analyses have suggested the existence of hierarchical modularity in the transcriptional regulatory network (TRN) of *Escherichia coli* K-12 [7-10]. Previous works have proposed methodologies from which this organization could be inferred [9-11]. These works suggested the existence of a pyramidal top-down hierarchy. Unfortunately, these approaches have proven inadequate for networks involving feedback loops (FBLs) or feedforward motifs (FFs) [10,11], two topological structures relevant to the organization and dynamics of TRNs [2,12-16]. In addition, module identification approaches frequently have been based on clustering methods, in which each gene must belong to a certain module [6,7,17]. Although analyses using these methods have reported good results, they have revealed two inconveniences: they rely on certain parameters or measurement criteria that, when modified, can generate different modules; and a network with scale-free properties foresees the existence of a small group of strongly connected nodes (hubs), but to what modules do these hubs belong? Maybe they do not belong to a particular module, but do they serve as coordinators of module responses?

Alternatively, we developed a novel algorithm to enumerate all the FBLs comprising two or more nodes existing in the TRN, thus providing the first systems-level enumeration and analysis of the global presence and participation of FBLs in the functional organization of a TRN. Our results show, contrary to what has been previously reported [9,10], the presence of positive and negative FBLs bridging different organizational levels of the TRN of *E. coli*. This new evidence highlights the necessity to develop a new strategy for inferring the hierarchical modular organization of TRNs.

To address these concerns, in this work we propose an alternative approach founded on inherent topological features of hierarchical modular networks. This approach recognizes hubs and classifies them as independent elements that do not possess a membership to any module, and reveals, in a natural way, the modules comprising the TRN by removing the hubs. This methodology enabled us to reveal the natural organization of the TRN of *E. coli*, where hierarchical transcription factors (hierarchical TFs) govern independent modules whose responses are integrated at the promoter level by intermodular genes.

## Results

The TRN of *E. coli* K-12 is the best characterized of all prokaryote organisms. In this work, the TRN was reconstructed using mainly data obtained from RegulonDB [18], complemented with new sigma factor interactions gathered from a literature review on transcriptional regulation mediated by sigma factors (see Materials and methods). In our graphical representation, each node represents a gene and each edge a regulatory interaction. The TRN used in this work was represented as a directed graph comprising 1,692 nodes (approximately 40% of the total genes in the genome) with 4,301 arcs (directed regulatory interactions) between them. Neglecting autoregulation and the directions of interactions between genes, the average shortest path of the network was 2.68, supporting the notion that the network has small-world properties [2]. The connectivity distribution of the TRN tends to follow a power law, $P(k) \sim k^{-2.06}$, which implies that it has scale-free properties (Figure S1a in Additional data file 1). In addition, the distribution of the clustering coefficient shows a power law behavior, with $C(k) \sim k^{-0.998}$ (Figure S1b in Additional data file 1). In the latter, the exponent value is virtually equal to -1, strongly suggesting that the network possesses a hierarchical modular architecture [2,19].

### The TRN has FBLs that involve mainly global and local TFs

The pioneering theoretical work of René Thomas [15,16,20,21] and experimental work [14,22] have shown the topological and dynamic relevance of feedback circuits (FBLs). In regulatory networks, FBLs are associated with biological phenomena, such as homeostasis, phenotypic variability, and differentiation [14,16,20,22]. Previous studies have established the importance of FBLs for both the modularity of regulatory networks [21] and their dynamics [14-16,20,22]. Ma *et al*. [9,10] suggested that FBLs that exist in the TRN of *E. coli* are not relevant for the topological organization of the TRN. Using an *E. coli* TRN reconstruction that included sigma factor interactions, they claimed to have identified only seven two-node FBLs (that is, FBLs with the structure A → B → A) and no FBLs comprising more than two nodes [10]. However, given that their approach requires, *a priori*, an acyclic network [23], genes involved in an FBL are placed in the same hierarchical layer, under the argument that they are in the same operon [10].

To get a global image of FBLs, an original algorithm was developed and implemented (see Materials and methods). This algorithm allowed us to enumerate all FBLs, comprising two or more nodes, existing in the TRN (Table 1). A total of 20

**Table 1**

**FBLs identified in the TRN of *Escherichia coli***

| Type of FBL | Number of genes | Genes | Interactions | Are genes in the same operon? |
| --- | --- | --- | --- | --- |
| + | 2 | *arcA fnr* | - - | No |
| - | 2 | *arcA fnr* | - + | No |
| - | 2 | *gadX hns* | + - | No |
| + | 2 | *gadX rpoS* | + + | No |
| - | 2 | *gutM srlR* | + - | Yes |
| - | 2 | *lexA rpoD* | - + | No |
| - | 2 | *marA marR* | + - | Yes |
| - | 2 | *marA rob* | - + | No |
| + | 2 | *rpoD rpoH* | + + | No |
| + | 3 | *crp rpoH rpoD* | + + + | No |
| - | 3 | *crp rpoH rpoD* | - + + | No |
| - | 3 | *cytR rpoH rpoD* | - + + | No |
| + | 3 | *gadE gadX rpoS* | + + + | No |
| + | 3 | *marA rob marR* | - + - | No |
| + | 3 | *rpoD rpoN rpoH* | + + + | No |
| - | 4 | *cpxR rpoE rpoH rpoD* | - + + + | No |
| - | 4 | *crp cytR rpoH rpoD* | + - + + | No |
| - | 5 | *IHF fis hns gadX rpoS* | + + - + + | No |
| - | 5 | *argP dnaA rpoH rpoD phoB* | + - + + + | No |
| - | 5 | *cpxR rpoE rpoN rpoH rpoD* | - + + + + | No |

Eighty percent of the total FBLs involve, at least, one global TF. The longest FBL comprises five TFs. Only two FBLs have genes encoded in the same operon, contrary to what was previously reported by Ma et al. [10], thus suggesting that these FBLs work as uncoupled systems. In addition, seven positive FBLs were identified, which potentially could give rise to multistability.

FBLs were found: 9 (45%) with two nodes and 11 (55%) with more than two nodes. It was found that FBLs in the TRN tend mainly to connect global TFs with local TFs (at this point we used the definitions of global and local TFs given by Martinez-Antonio and Collado-Vides [24]). It was also found that only 2 FBLs (10%) are located in the same operon, 4 (20%) involve only local TFs, 10 (50%) involve both global and local TFs, and 6 (30%) involve only global TFs. We observed a couple of dual FBLs, the first comprising *arcA* and *fnr* and the second comprising *crp*, *rpoH*, and *rpoD*. These dual FBLs comprise dual regulatory interactions, thus giving rise to two overlapping FBLs, one positive and the other negative. However, each of these overlapping FBLs was enumerated as a different FBL, given that the dynamic behaviors of positive and negative FBLs are quite different.

### Nodes of hierarchical modular networks can be classified into one of two possible classes: hierarchical or modular nodes

The characteristic signature of hierarchical modularity in a network is the clustering coefficient distribution, which must follow a power law, $C(k) \sim k^{-1}$ [2,19]. This coefficient measures how much the nearest neighbors of a TF affect each other, thus providing a measure of the modularity for the TF. In the extreme limits of the clustering coefficient distribution, nodes follow two apparently contradictory behaviors [2] (Figure 1a). At low connectivity, nodes show high clustering coefficients. On the contrary, at high connectivity, nodes show low clustering coefficients. Previous work with the *E. coli* metabolic network [17] suggested that the first behavior is due to network modularity but the latter is due to the presence of hubs. In addition, a previous analysis of the TRN of *Saccharomyces cerevisiae* found that direct connections between hubs tend to be suppressed while connections between hubs and poorly connected nodes are favored [25], suggesting that modules tend to be organized around hubs. This evidence suggested two possible roles for nodes: nodes that shape modules (they have low connectivity and a high clustering coefficient, which will be called modular nodes); and nodes that bridge modules (they have high connectivity and a low clustering coefficient, which will be called hierarchical nodes), establishing in this way a hierarchy that dynamically governs module responses.

It can be observed in $C(k)$ distributions following a power law that initially slight increments in the connectivity value ($k$) will make the clustering coefficient decrease quickly. However, eventually a point is reached where the situation is inverted. Then, a larger increment in connectivity is needed to make the clustering coefficient decrease. From this behavior the existence of an equilibrium point in the $C(k)$ distribution is inferred, where the variation of the clustering

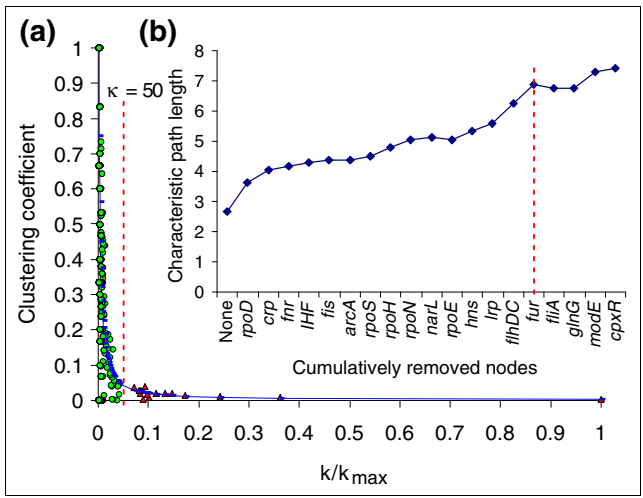

**Figure 1**
Identification of hierarchical and modular nodes. **(a)** Distribution of the clustering coefficient, $C(k)$, and calculated κ value. The blue line represents the $C(k)$ power law. The dashed red line indicates the κ value obtained for this $C(k)$ distribution. Red triangles represent hierarchical nodes, while green circles indicate modular nodes. **(b)** The characteristic path length after cumulative removal of all hierarchical nodes and some modular ones. The red dashed line indicates the sudden change in the original increasing tendency when the last hierarchical TFs (FlhDC and Fur) were removed. This suggests that the removal of hierarchical nodes broke the connections bridging modules, thus disintegrating the TRN.

coefficient is equal to the variation of connectivity but with the opposite sign:

$$dC(k)/dk = -1$$

Solving this equation gives the connectivity value (κ) where such an equilibrium is reached (see Material and methods). Herein, κ is proposed as a cutoff value that disaggregates the set of nodes into two classes (Figure 1a). Hierarchical nodes are those with connectivity greater than κ. On the other hand, modular nodes are those with connectivity less than κ.

The κ value can be calculated with the formula (see Materials and methods):

$$\kappa = \sqrt[\alpha+1]{\alpha\gamma} \cdot k_{max}$$

This formula relates the equilibrium point (κ) of the $C(k)$ distribution with its exponent ($-\alpha$) and its proportionality constant (γ). It has been shown that in 'ideal' hierarchical modular networks the exponent $-\alpha$ is equal to -1 [2,19]. Thus, substituting this value into the previous formula gives:

$$\kappa = \sqrt{\gamma} \cdot k_{max}$$

Therefore, in 'ideal' networks the equilibrium point depends exclusively on the proportionality constant of $C(k)$. To the best of our knowledge, this is the first time that a relevant topological interpretation has been given to the proportionality constant.

### Hierarchical nodes correlate highly with known global TFs

After computing the κ value for the TRN, the following 15 TFs were identified as hierarchical nodes (nodes with connectivity greater than 50; Figure 1): RpoD ($\sigma^{70}$), CRP, FNR, IHF, Fis, ArcA, RpoS ($\sigma^{38}$), RpoH ($\sigma^{32}$), RpoN ($\sigma^{54}$), NarL, RpoE ($\sigma^{24}$), H-NS, Lrp, FlhDC, and Fur. All these TFs, except FlhDC and Fur, have been reported several times as global TFs [13,24,26,27]. In addition, Madan Babu and Teichmann [27] have previously reported Fur as a global TF. FlhDC and Fur regulate genes with several physiological functions, which makes them potential candidates to be global TFs [28]. Fur regulates amino acid biosynthesis genes [29], $Fe^+$ transport [30-32], flagellum biosynthesis [29], the Krebs cycle [33], and Fe-S cluster assembly [34]. On the other hand, FlhDC mainly regulates membrane genes. Nevertheless, these genes take part in several physiological functions, such as motility [35], glutamate [36] and galactose [37] transport, anaerobiosis [37], and 3-P-glycerate degradation [37]. When connectivity was less than κ, genes encoding local TFs (herein called modular TFs) and structural genes were found. FliA ($\sigma^{28}$) and FecI ($\sigma^{19}$) sigma factors are in the group of modular nodes. This is understandable, because both respond to very specific cell conditions (flagellum biosynthesis and citrate-dependent $Fe^+$ transport, respectively), and they affect the transcription of few genes (43 and 6 genes, respectively). These results suggest that the κ value may be a good predictor for global TFs.

### Hierarchical nodes act as bridges keeping modules connected

The characteristic path length is defined as the average of the shortest paths between all pairs of nodes in a network. It is a measure of the global connectivity of the network [38]. Using an *in silico* strategy, the effect on the characteristic path length when attacking hierarchical nodes was analyzed. In order to do this, all hierarchical nodes and some modular ones were removed one by one in decreasing order of connectivity (Figure 1b). The removal of hierarchical nodes increased, following a linear tendency, the characteristic path length from 2.7 to 6.9. However, when the last two hierarchical nodes (*flhDC* and *fur*) were removed, a sudden change was observed in the tendency, followed by a stabilization when some modular nodes were removed, therefore supporting the idea that removal of hierarchical nodes disintegrates the TRN by breaking the bridges that keep modules together.

### Identification of modules in the TRN

The removal of hierarchical nodes revealed 62 subnetworks or modules (see Materials and methods; Additional data file 2) and left 691 isolated genes. An analysis of the biological function of the isolated genes showed that many of them are elements of the basal machinery of the cell (tRNAs and its charging enzymes, DNA and RNA polymerases, ribosomal proteins and RNAs, enzymes of the tricarboxylic acid cycle and respiratory chain, DNA methylation enzymes, and so on). The regulation of these genes, whose products must be constantly present in the cell, is mediated only by hierarchical TFs. One of the identified modules (module 5) comprises 606 genes (35% of the analyzed TRN). This megamodule suggested the existence of other elements, in addition to hierarchical nodes, that connect modules. We know that a TRN that has been reconstructed while neglecting structural genes does not show the existence of a megamodule (JAF-G, unpublished data). Therefore, an intermodular gene was defined as a structural gene whose expression is modulated by TFs belonging to two or more submodules. To identify these intermodular genes, the megamodule was isolated and structural genes removed. This revealed the submodule cores (islands of modular TFs) shaping the megamodule (see Materials and methods). The megamodule comprises 39 submodules connected by the regulation of 136 intermodular genes, which are organized into approximately 55 transcriptional units (Additional data file 3).

To determine the biological relevance of the theoretically identified modules, two independent analyses were performed. On the one hand, one of us (LGT-Q) used biological knowledge to perform a manual annotation of identified modules. On the other hand, two of us (JAF-G and JAA-P) made a blind-automated annotation based on functional class, according to the MultiFun system [39], that showed a statistically significant enrichment ($p$-value <0.05; see Materials and methods). Both analyses showed similar conclusions. The blind-automated method found that 97% of modules show enrichment in terms of functional classes. However, it was observed that the manual analysis added subtle details that were not evident in the automated analysis due to incompleteness in the MultiFun system (Additional data file 2). At the module level, it was found that *E. coli* mainly has systems for carbon source catabolism, cellular stress response, and ion homeostasis. In addition, it was found that the 39 submodules comprising the megamodule could be grouped according to their biological functions into seven regions interconnected by intermodular genes (Figure 2). The most interconnected regions involve nitrogen and sulfur assimilation, carbon source catabolism, cellular stress response, respiration forms, and oxidative stress.

### Inference of the hierarchy governing the TRN

For more than 20 years it has been recognized that regulatory networks comprise complex circuits with different control levels. This makes them able to control different subroutines of the genetic program simultaneously [28,40]. Recently, global topological analyses have suggested the existence of hierarchical modularity in TRNs [2,7,8]. Previous works proposed methodologies to infer this hierarchical modular organization [9-11]. Unfortunately, the previous methodological approaches have been shown to be inadequate to deal with FFs and FBLs [10,11], two relevant topological structures. On the other hand, biological conclusions obtained with these approaches were counterintuitive, as they placed, in the highest hierarchical layers, TFs that respond to very specific conditions of the cell and which, therefore, lack pleiotropic effects.

Gottesman [28] defined a global TF as one that: regulates many genes; entails regulated genes that participate in more than one metabolic pathway; and coordinates the expression of a group of genes when responding to a common need (for detailed definitions of global and local TFs please refer to the work of Martinez-Antonio and Collado-Vides [24]). Based on Gottesman's ideas, it could be asked if a modular organization requires a hierarchy to coordinate module responses. To address this concern, based on the definition proposed by Gottesman and using the concept of pleiotropy, a methodology to infer the hierarchy governing the TRN was developed. For this methodology, nodes belonging to the same module were shrunk into a single node, and a bottom-up approach was used (see Materials and methods). This approach places each hierarchical TF in a specific layer, depending on two factors: theoretical pleiotropy (the number of regulated modules and hierarchical TFs); and the presence of direct regulation over hierarchical TFs placed in the immediate lower hierarchical layer. This second factor was taken into account because a hierarchical TF may indirectly propagate its control to other modules, by changing the expression pattern of a second hierarchical TF that directly controls them. Given that a hierarchical layer does not depend on the number of genes regulated by a hierarchical TF, but on the number of modules, it is worth mentioning that this approach is not based on connectivity. Therefore, given that each module is in charge of a different physiological response, it can be argued that this approach is founded on pleiotropy.

Five global chains of command were found, showing the regulatory interactions between hierarchical TFs (Figure 3). Each of the chains of command is in charge of global functions in the cell. In addition, in the highest hierarchical layers, the presence of six hierarchical TFs was observed, three of them (RpoD, CRP, and FNR) governing more than one of these global chains of command. The expression of IHF, in spite of the fact that it only governs one global chain of command, can be affected by a different chain from a lower hierarchy (RpoS) [41]. Each of these TFs sends signals of general interest to a large number of genes in the cell. RpoD ($\sigma^{70}$) is the housekeeping sigma factor, and it can indicate to the cellular machinery the growth phase of the cell or the lack of any stress [42]. CRP-cAMP alerts the cell to low levels of energy uptake, allowing a metabolic response [43]. IHF (besides Fis and H-NS) senses DNA supercoiling, thus indirectly sensing many environmental conditions (growth phase, energy level, osmolarity, temperature, pH, and so on) that affect this DNA property [44]. This supports the idea that DNA supercoiling itself might act as a principal coordinator of global gene expression [45,46]. Finally, FNR senses extracellular oxygen levels, permitting, through coregulation with ArcA and NarL, a proper respiratory response [47,48]. RpoN, with $\sigma^{54}$-dependent activators, controls gene expression to coordinate nitrogen assimilation [49]. RpoE ($\sigma^{24}$) reacts to stress signals outside the cytoplasmic membrane by transcriptional activation of genes encoding products involved in membrane protection or repair [50].

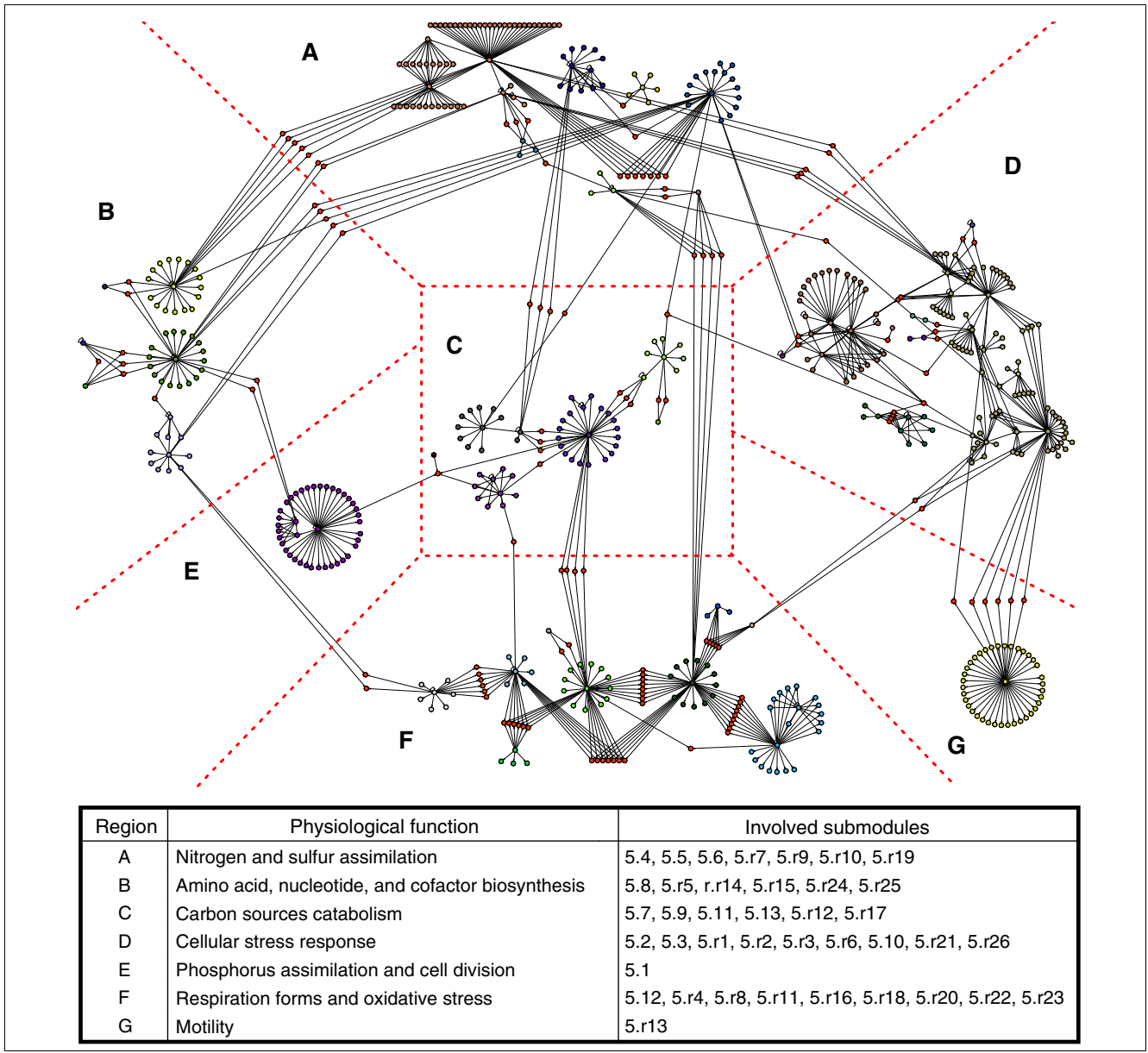

**Figure 2**
Empirical grouping, into seven regions, of submodules comprising the megamodule. Each color represents a submodule, while intermodular genes are shown in orange. Intermodular genes are placed inside the region that best associates with its most important physiological function. For example, the intermodular gene *amtB*, positively regulated by NtrC (region A) and GadX (region D), encodes an ammonium transporter under acidic growing conditions. Therefore, this gene was placed in the nitrogen and sulfur assimilation region (region A).

### FFs mainly bridge modules shaping the TRN hierarchical backbone

A remarkable feature of complex networks is the existence of topological motifs [12,13]. It has been previously suggested that they constitute the building blocks of complex networks [8,12]. Nevertheless, recent studies have provided evidence that overabundance of motifs does not have a functional or evolutionary counterpart [51-54]. Indeed, some studies have suggested that motifs could be by-products of biological network organization and evolution [52,53,55]. In particular, work by Ingram *et al.* [54] has shown that the bi-fan motif can exhibit a wide range of dynamic behaviors. Given that, we concentrated our analysis on three-node motifs.

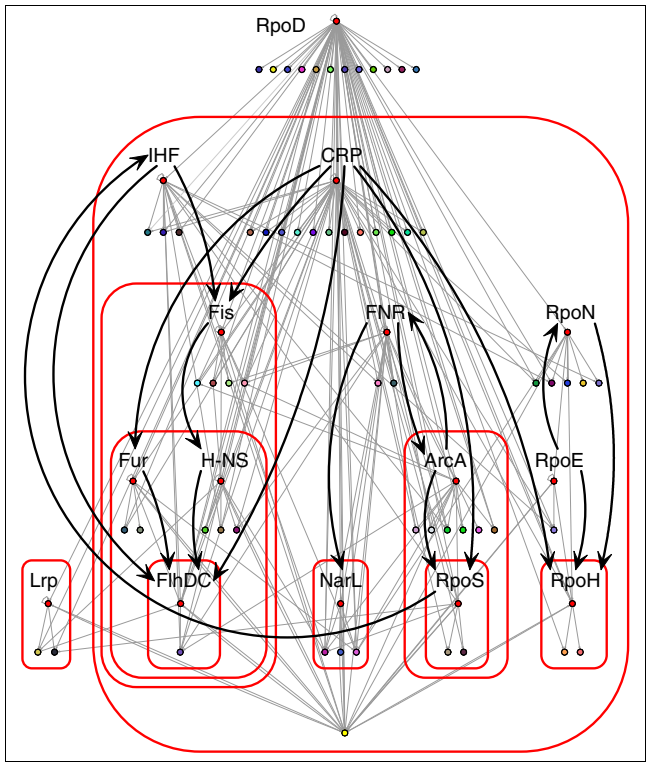

**Figure 3**
Hierarchical modular organization map of subroutines comprising the genetic program in *E. coli*. Each color represents a module, while hierarchical TFs are shown in red. Black arrows indicate the regulatory interactions between hierarchical TFs. For the sake of clarity, RpoD interactions are not shown, and the megamodule is shown as a single yellow node at the bottom. However, according to our data, RpoD affects the transcription of all hierarchical TFs, except RpoE, while RpoD, RpoH, and LexA (a modular TF) could affect RpoD expression. Red rounded-corner rectangles bound hierarchical layers. The presence of five global chains of command is noted: host/free-life sensor and type 1 fimbriae (Lrp); replication, recombination, pili, and extracytoplasmic elements (Fis, Fur, H-NS, FlhDC); respiration forms (NarL); starvation stress (ArcA, RpoS); and heat shock (RpoH). Lrp appears disconnected from other hierarchical TFs because, to date, it is only known that RpoD, Lrp, and GadE (a modular TF) modulate its expression.

We identified the entire repertoire of three-node network motifs present in the *E. coli* TRN by using the mfinder program [12]. Thus, we identified two three-node network motifs: the FF; and an alternative version of an FF merging an FBL between the regulatory nodes. It suggests that the FF is the fundamental three-node motif in the *E. coli* TRN. In order to analyze FF participation in the hierarchy inferred by our methodology, the effect of the removal of hierarchical nodes on the total number of FFs in the TRN was analyzed (Figure 4a). The fraction of remaining FFs after cumulative removal of hierarchical nodes, in decreasing connectivity order, was computed. It was found that the sole removal of *rpoD* ($\sigma^{70}$) and *crp*, the two most-connected hierarchical nodes in the TRN, decreased to 22% the total FFs. However, the removal of all hierarchical nodes decreased the total FFs to 3.5%, in agreement with previous work suggesting that FFs tend to cluster around hubs [56]. Our results showed that 96.5% of the total FFs are in the TRN bridge modules, while the remaining 3.5% are within modules. This evidence suggests that the FF role is to bridge modules, shaping a hierarchical structure governed by hierarchical TFs.

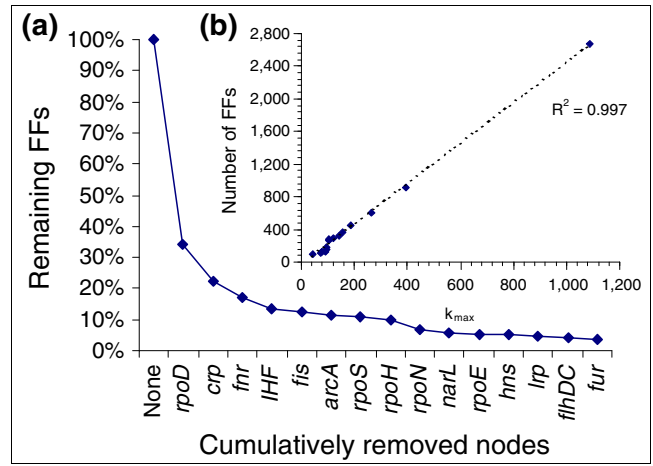

**Figure 4**
FFs bridge modules and shape the backbone of the hierarchy governing the TRN. **(a)** Remaining TFs after cumulative removal of hierarchical nodes. The removal of all hierarchical nodes decreased to 3.5% the total FFs. **(b)** Correlation between FF number and maximum connectivity for each attacked network. The FF number is proportional to the number of links of the most-connected hierarchical node, thus suggesting that FFs are the backbone of the hierarchy in the TRN.

The correlation between FF number and maximum connectivity (number of links of the most-connected node, $k_{max}$) for each attacked network was analyzed (Figure 4b). It was found that the FF number linearly correlated with the maximum connectivity. As hierarchical nodes were removed, the FF number decreased proportionally with the maximum connectivity of the corresponding attacked network. All this shows that hierarchical TFs are intrinsically related to FFs, suggesting that, in addition to bridging modules, FFs are the backbone of the hierarchical organization of the TRN.

## Discussion
Contrary to what has been previously reported [9,10], we found FBLs involving different hierarchical layers, which implies that the expression of some hierarchical TFs also may depend on modular TFs, thus allowing the reconfiguration of the regulatory machinery in response to the fine environmental sensing performed, through allosterism, by modular TFs. On the other hand, a network with FBLs poses a paradox when inferring its hierarchy. Given the circular nature of interactions, what nodes should be placed in a higher hierar-chical layer? This paradox was solved using the κ value to identify hierarchical and modular elements and then using the theoretical pleiotropy to infer the hierarchy governing the TRN.

Global TFs have been proposed using diverse relative measures [9,10,13,24,27,28]; unfortunately, currently there is not a consensus on the best criteria to identify them. Gottesman's seminal paper [28] was the first to define the properties for which a TF should be considered a global TF. Martinez-Antonio and Collado-Vides [24] conducted a review and analyzed several properties, searching for diagnostic criteria to identify global TFs. Nevertheless, while these authors did shed light on relevant properties that could contribute to identification of global TFs, they did not reach any explicit diagnostic criteria. The κ value showed high predictive power, as all known global TFs were identified, and even more, the existence of two new global TFs is proposed: FlhDC and Fur. Recently, an analysis of the TRN of *Bacillus subtilis* supported the predictive ability of this method (JAF-G, unpublished data), offering the possible first mathematical criterion to identify global TFs in a cell. This criterion allowed us to show that, in spite of its apparent complexity, the TRN of *E. coli* possesses a singular elegance in the organization of its genetic program. Only 15 hierarchical TFs (0.89% of the total nodes) coordinate the response of the 100 identified modules (50.23% of the total nodes). All the modules identified by Resendis-Antonio *et al.* [7] were recovered by our methodology. However, given that in this study the TRN includes structural genes, we could identify 87 new modules. Therefore, our approach allows fine-grain identification of modules, for example, modules responsible for catabolism of specific carbon sources. There are 691 genes (40.84% of the total nodes) that mainly encode cellular basal elements. The existence of one megamodule led us to define intermodular genes and to identify 136 of them (8.04% of the total nodes). It was found that submodules with similar functions tend to agglomerate into seven regions, thus shaping the megamodule. Therefore, at a TRN level, data processing follows independent casual chains for each module, which are globally governed by hierarchical TFs. Thus, hierarchical TFs coordinate the cellular system responses as a whole by letting modules get ready to react in response to external stimuli of common interest, while modules retain their independence, responding to stimuli of local interest. On the other hand, intermodular genes integrate, at the promoter level, the incoming signals from different modules. These promoters act as molecular multiplexers, integrating different physiological signals in order to make complex decisions. Examples of this are the *aceBAK* and *carAB* operons. The *aceBAK* operon encodes glyoxylate shunt enzymes. The expression of this operon is modulated by FruR [57] (module 5.11, gluconeogenesis) and IclR [58] (module 5.13, aerobic fatty acid oxidation pathway). This operon could integrate the responses of these two modules in order to keep the balance between energy production from fatty acid oxidation and gluconeogenesis activation for biosynthesis of building blocks.

On the other hand, the *carAB* operon encodes a carbamoyl phosphate synthetase. The expression of this operon is controlled by PurR [59] (module 5.r25, purine and pyrimidine biosynthesis), ArgR [60] (module 5.r5, L-ornithine and L-arginine biosynthesis), and PepA [59] (5.r24, carbamoyl phosphate biosynthesis and aminopeptidase A/I regulation). This is an example where different modules could work as coordinators of a shared resource. The promoter of this operon could integrate the responses of the modules to coordinate the expression of an enzyme whose product, carbamoyl phosphate, is a common intermediary for the *de novo* biosynthesis of pyrimidines and arginine. This evidence shows a novel nonpyramidal architecture in which independent modules are globally governed by hierarchical transcription factors while module responses are integrated at the promoter level by intermodular genes.

The clustering coefficient is a strong indicator of modularity in a network. It also quantifies the presence of triangular substructures. The TRN shows a high average clustering coefficient, implying a high amount of triangular substructures. Indeed, the probability of a node being a common vertex of $n$ triangles decreases as the number of involved triangles increases, following the power law $T(n) \sim n^{-1.95}$ (Figure S1c in Additional data file 1). In other words, if a node is arbitrarily chosen, the probability of it being the vertex of a few triangles is high. This also implies that many triangles have as a common vertex a small group of nodes. On the other hand, in a directed graph there are only two basic triangular substructures: FFs and three-node FBLs. By merging two-node FBLs with these two triangular substructures, it is possible to create variations of them. It was found that the number of two-node and three-node FBLs (eight and five FBLs, respectively) was much lower than the total number of FFs (2,674 FFs). These results imply that triangular substructures are mainly FFs or variations of them. Besides, FFs mainly comprise, at least, one hierarchical node [56] (Figure 4). This is in agreement with the observation that many triangles possess as a common vertex a small group of nodes. Here it was shown that hierarchical nodes and their interactions shape the backbone of the TRN hierarchy. Therefore, FFs are strongly involved in the hierarchical modular organization of the TRN of *E. coli*, where they act as bridges connecting genes with diverse physiological functions. Resendis-Antonio *et al.* [7] showed that FFs are mainly located within modules. Nevertheless, given that in this study it was determined that hubs do not belong to modules, it was found that FFs shape the hierarchy of the TRN bridging modules in a hierarchical fashion. This supports the findings of Mazurie *et al.* [52], showing that FFs are a consequence of the network organization and they are not involved in specific physiological functions.

## Conclusions

The study of the topological organization of biological networks is still an interesting research topic. Methodologies for node classification and natural decomposition, such as the one proposed herein, allow identification of key components of a biological network. This approach also enables the analysis of complex networks by using a zoomable map approach, helping us understand how their components are organized in a meaningful way. In addition, component classification could shed light on how different networks (transcriptional, metabolic, protein-protein, and so on) interface with each other, thus providing an integral understanding of cellular processes. The herein-proposed approach has promising applications for unraveling the functional architecture of the TRNs of other organisms, allowing us to gain a better understanding of their key elements and their interrelationships. In addition, it provides a large set of experimentally testable hypotheses, from novel FBLs to intermodular genes, which could be a useful guide for experimentalists in the systems biology field. Finally, network decomposition into modules with well-defined inputs and outputs, and the suggestion that they process information in independent casual chains governed by hierarchical TFs, would eventually help in the isolation, and subsequent modeling, of different cellular processes.

## Materials and methods
### Data extraction and TRN reconstruction

To reconstruct the TRN, structural genes, sigma factor-encoding genes, and regulatory protein-encoding genes were included (the full data set is available as Additional data file 4). Two flat files with data (NetWorkSet.txt and SigmaNetWorkSet.txt) were downloaded from RegulonDB version 5.0 [18,61]. From the NetWorkSet.txt file, 3,001 interactions between regulatory proteins and regulated genes were obtained. From the SigmaNetWorkSet.txt file, 1,488 interactions between sigma factors and their transcribed genes were obtained. Next, this information was complemented with 81 new interactions found in a literature review of transcribed promoters by the seven known sigma factors of *E. coli* (these interactions account for 5.4% of the total sigma factor interactions in the reconstructed TRN and currently are integrated and available in RegulonDB version 6.1). The criteria used to gather the additional sigma factor interactions from the literature were the same as those used by the RegulonDB team of curators. In our graphic model, sigma factors were included as activator TFs because their presence is a necessary condition for transcription to occur. Indeed, some works [62-64] have shown that there are TFs that are able to interact with free polymerase before binding to a promoter, in a way reminiscent of the mechanism used by sigma factors. To avoid duplicated interactions, heteromeric TFs (for example, IHF encoded by *ihfA* and *ihfB* genes, HU encoded by *hupA* and *hupB*, FlhDC encoded by *flhC* and *flhD*, and GatR encoded by *gatR_1* and *gatR_2*) were represented as only one node, given that there is no evidence indicating that any of the subunits have regulatory activity *per se*.

### Software
For the analysis and graphic display of the TRN, Cytoscape [65] was used. To identify FFs, the mfinder program [12] was used. To calculate κ values, computational annotations, and other numeric and informatics tasks, Microsoft Excel and Microsoft Access were used.

### Algorithm for FBL enumeration
First, The TRN was represented, neglecting autoregulation, as a matrix of signs (**S**). Thus, each $S_{i,j}$ element could take a value in the set **{+,-,D,0}**, where '+' means that $i$ activates $j$ transcription, '-' means than $i$ represses $j$ transcription, **D** means that $i$ has a dual effect (both activator and repressor) over $j$, and **0** means that there is no interaction between $i$ and $j$. Second, All nodes with incoming connectivity or outgoing connectivity equal to zero were removed. Third, the transitive closure matrix of the TRN (**M**) was computed using a modified version of the Floyd-Warshall algorithm [23]. Each $M_{i,j}$ element could take a value in the set **{0,1}**, where **0** means that there is no path between $i$ and $j$ and **1** means that, at least, there is one path between $i$ and $j$. Fourth, for each $M_{i,i}$ element equal to **1**, a depth-first search beginning at node $i$ was done, marking each visited node. The depth-first search stopping criterion relies on two conditions: first, when node $i$ is visited again, that is, an FBL ($i \to ... \to i$) is identified; second, when a previously visited node, different from $i$, is visited again. Fifth, isomorphic subgraphs were discarded from identified FBLs.

### κ value calculation
For each node in the TRN, connectivity (as a fraction of maximum connectivity, $k_{max}$) and the clustering coefficient were calculated. Next, the $C(k)$ distribution was obtained using least-squares fitting. Given $C(k) = \gamma k^{-\alpha}$, the equation:

$$dC(k)/dk = -1$$

has as its solution the formula:

$$\kappa = \sqrt[\alpha+1]{\alpha \gamma} \cdot k_{max}.$$

### Module identification
The algorithm to identify modules used a natural decomposition approach. First, the κ value was calculated for the TRN of *E. coli*, yielding the value of 50. Then, all hierarchical nodes (nodes with $k > \kappa$) were removed from the network. Therefore, the TRN breaks up into isolated islands, each comprising interconnected nodes. Finally, each island was considered a module.

### Identification of submodules and intermodular genes comprising the megamodule
The megamodule was isolated and all structural genes were removed, breaking it up into isolated islands. Next, each island was identified as a submodule. Finally, all the removed structural genes and their interactions were added to the net-work according to the following rule: if a structural gene G is regulated only by TFs belonging to submodule M, then gene G was added to submodule M. On the contrary, if gene G is regulated by TFs belonging to two or more submodules, then gene G was classified as an intermodular gene.

### Manual annotation of identified modules
Manual annotation of physiological functions of identified modules was done using the biological information available in RegulonDB [18,61] and EcoCyc [66,67].

### Computational annotation of identified modules
Each gene was annotated with its corresponding functional class according to Monica Riley's MultiFun system, available via the GeneProtEC database [39,68]. Next, *p*-values, as a measure of randomness in functional class distributions through identified modules, were computed based on the following hypergeometric distribution: let $N = 1,692$ be the total number of genes in the TRN and $A$ the number of these genes with a particular $F$ annotation; the *p*-value is defined as the probability of observing, at least, $x$ genes with an $F$ annotation in a module with $n$ genes. This *p*-value is determined with the following formula:

$$p\text{-value} = \sum_{i=x}^{n} \frac{\binom{A}{i}\binom{N-A}{n-i}}{\binom{N}{n}}.$$

Thus, for each module, the *p*-value of each functional assignment present in the module was computed. The functional assignment of the module was the one that showed the lowest *p*-value, if and only if it was less than 0.05.

### Inference of the hierarchy
To infer the hierarchy, a shrunken network was used, where each node represents a module or a hierarchical element. Hierarchical layers were created following a bottom-up approach and considering the number of regulated elements (theoretical pleiotropy) by hierarchical nodes, neglecting autoregulation, as follows. First, all nodes belonging to the same module were shrunk into a single node. Second, for each hierarchical element, the theoretical pleiotropy was computed. Third, the hierarchical element with lower theoretical pleiotropy and its regulated modules were placed in the lower hierarchical layer. Fourth, each hierarchical element and its regulated modules were added one by one in order of increasing theoretical pleiotropy. Fifth, if the added hierarchical element regulated, at least, one hierarchical element in the immediate lower layer, a new hierarchical layer was created; otherwise, the hierarchical element was added to the same hierarchical layer.

### Abbreviations
FBL, feedback loop; FF, feedforward topological motif; TF, transcription factor; TRN, transcriptional regulatory network.

### Authors' contributions
JAF-G and JC-V designed the research; JAF-G conceived the approach and designed algorithms; JAA-P and LGT-Q contributed to the algorithm to infer hierarchy; JC-V proposed the computational annotation of modules; JAF-G, JAA-P, and LGT-Q performed research; JAF-G, JAA-P, and LGT-Q contributed analytic tools; JAF-G, JAA-P, and LGT-Q analyzed data; JAF-G, JAA-P, LGT-Q, and JC-V wrote the paper.

### Additional data files
The following additional data are available. Additional data file 1 contains the topological properties of the transcriptional regulatory network of *E. coli*. Additional data file 2 is a table listing all the modules identified in this study and their manual and computational annotations. Additional data file 3 contains a listing of all the intermodular genes found in this study, their biological descriptions and roles as integrative elements. Additional data file 4 is a flat file with the full data set for the *E. coli* transcriptional regulatory network reconstructed for our analyses as described in the Materials and methods section.

### Acknowledgements
We thank Veronika E Rohen for critical reading of the statistical methodology used for the computational annotation of modules. We thank Mario Sandoval for help in codifying the algorithm for FBL enumeration. We also thank Patricia Romero for technical support. JAF-G was supported by PhD fellowship 176341 from CONACyT-México and was a recipient of a graduate complementary fellowship from DGEP-UNAM. This work was partially supported by grants 47609-A from CONACyT, IN214905 from PAPIIT-UNAM, and NIH RO1 GM071962-04 to JC-V.